\documentclass[12pt,preprint]{aastex}

\received{}
\accepted{}

\shorttitle{Pulsar Geodetic Spin Precession}
\shortauthors{Weisberg \& Taylor}

\begin{document}

\title{General Relativistic Geodetic Spin Precession in Binary Pulsar
B1913+16:  Mapping the Emission Beam in Two Dimensions}

\author{Joel M. Weisberg\altaffilmark{1}}
\affil{Department of Physics and Astronomy, Carleton College,
    Northfield, MN 55057}

\and

\author{Joseph H. Taylor\altaffilmark{2}}
\affil{Department of Physics, Princeton University,
Princeton, NJ 08544 USA}

\email{jweisber@carleton.edu}
\email{joe@pulsar.princeton.edu}

\altaffiltext{1}{Email:  {\it{jweisber@carleton.edu}}}

\altaffiltext{2}{Email: {\it{joe@pulsar.princeton.edu}}}

\slugcomment{Submitted to ApJ February 2002. Accepted May 2002}

\begin{abstract}
We have carefully measured the pulse profile of the binary pulsar PSR B1913+16
at 21 cm wavelength for twenty years, in order to search for variations
that result from general relativistic geodetic precession of the spin axis.
The profile width is found to decrease with time in its inner regions, while
staying essentially constant on its outer skirts.  We fit these data to a model
of the beam shape and precession geometry.  Four equivalent solutions
are found, but evolutionary considerations and polarization data select a 
single preferred model.  While the current data sample only a limited range
of latitudes owing to the long precessional cycle, the preferred model shows 
a beam elongated in the latitude direction and hourglass--shaped.

\end{abstract}


\section{Introduction}

The binary pulsar PSR B1913+16 resides in the strong gravitational field of 
its companion.  As a result, a wide variety of relativistic phenomena
are observable, including advance of periastron, gravitational redshift, and,
most notably, orbital decay due to gravitational radiation emission (Taylor
\& Weisberg 1989; Taylor et al. 1992).  Here we report on an additional 
general relativistic phenomenon, the   gravitomagnetic 
spin--orbital coupling known as geodetic precession.  This precession
provides a unique opportunity to study the pulsar beam along the latitude  
direction, orthogonal to the usual longitude dimension revealed by  pulsars' 
spinning across our line of sight.

Geodetic precession and resulting pulseshape changes in PSR B1913+16  were 
predicted shortly after the pulsar's discovery in 1974 (Damour \& Ruffini 
1974;
Barker \& O'Connell 1975a,b; Esposito \& Harrison 1975). The expected
signature of this phenomenon would be a change in the longitude separation of 
the two
main pulse components, which are thought to be  emitted from opposite sides 
of a hollow cone, as the line of sight slowly precesses across the cone.  The 
first observational searches for evidence of geodetic precession were 
reported by Weisberg, Romani, \& Taylor (1989) and Weisberg \& Taylor (1992).
The expected component separation variation was {\em{not}} observed, although
a secular change in the relative   {\em{intensity}} of the two main components
was detected. These observations were interpreted by the authors as resulting
from our line of sight fortuitously precessing across the  {\em{middle}} (to
explain the constancy of component separation) of a patchy (to justify the
relative intensity variation) emission beam.  \citet{I91} then predicted on 
the basis of these observations
that the beam will precess out of our line of sight in the year $\sim 2020$.
The expected component separation change was finally detected by 
\citet{Kramer98}, using data from the Effelsberg 100--m dish combined 
with our earlier results.  Kramer used data on the separation between 
the two profile peaks as a function of time to fit a model for the 
geometry of the precession
and of the beam.  Here we augment our earlier measurements with new Arecibo
post--upgrading observations over the last few years.  We measure the
beamwidth as a function of time at fourteen different intensity levels, 
and we fit these data to a model of the beam geometry to arrive at a 
picture of the two--dimensional structure of the emission region.
 
\section{Data Acquisition}

All data were gathered at Arecibo Observatory at frequencies near 1400 MHz 
with one of a series of dedicated pulsar backends,
called the Mark I, II, and III systems [see \citet{TW89} for details].  
In all cases, two orthogonal
circularly polarized signals were sampled, detected, dedispersed, folded 
at the apparent 
pulsar period, and integrated on--line for five minutes.  The resulting 
total power pulse profile was stored to tape as the basic processing unit, a
``scan.''

The scans from a given observing session, typically consisting of fourteen 
2--3 h daily blocks of data acquisition, were then summed to create a
``session--average'' profile.  All subsequent analyses used these twenty--two 
session average profiles, acquired between 1981 and 2001.

In order to make the Mark I, II, and III data as closely intercomparable as
possible, we  simulated in software some of the on-line hardware.
We chose to smooth the Mark I and II data to match the coarser 
resolution of the Mark III system, with its $312 \, \mu s$ exponential time
constant, 1.25 MHz individual frequency channels, and 512 phase bins. (The 
Mark 
I and Mark II data had already been made effectively equivalent by 
convolving 
the Mark II data with a $100 \, \mu s$ time constant\citep{Weisberg89}. 
The Mark I and equivalent Mark II data were convolved with a decaying 
exponential function of $300 \, \mu s$ time constant to match the Mark 
III hardware exponential time constant, and with an 11-bin ($650 
\, \mu s$) boxcar function to simulate the dispersion smearing across the 
individual Mark III filter channels. Adjacent pairs of the resulting 
1024--bin waveform were combined to create final  512--bin Mark I and 
Mark II profiles 
equivalent to the Mark III profiles.

\begin{figure*}
\plotone{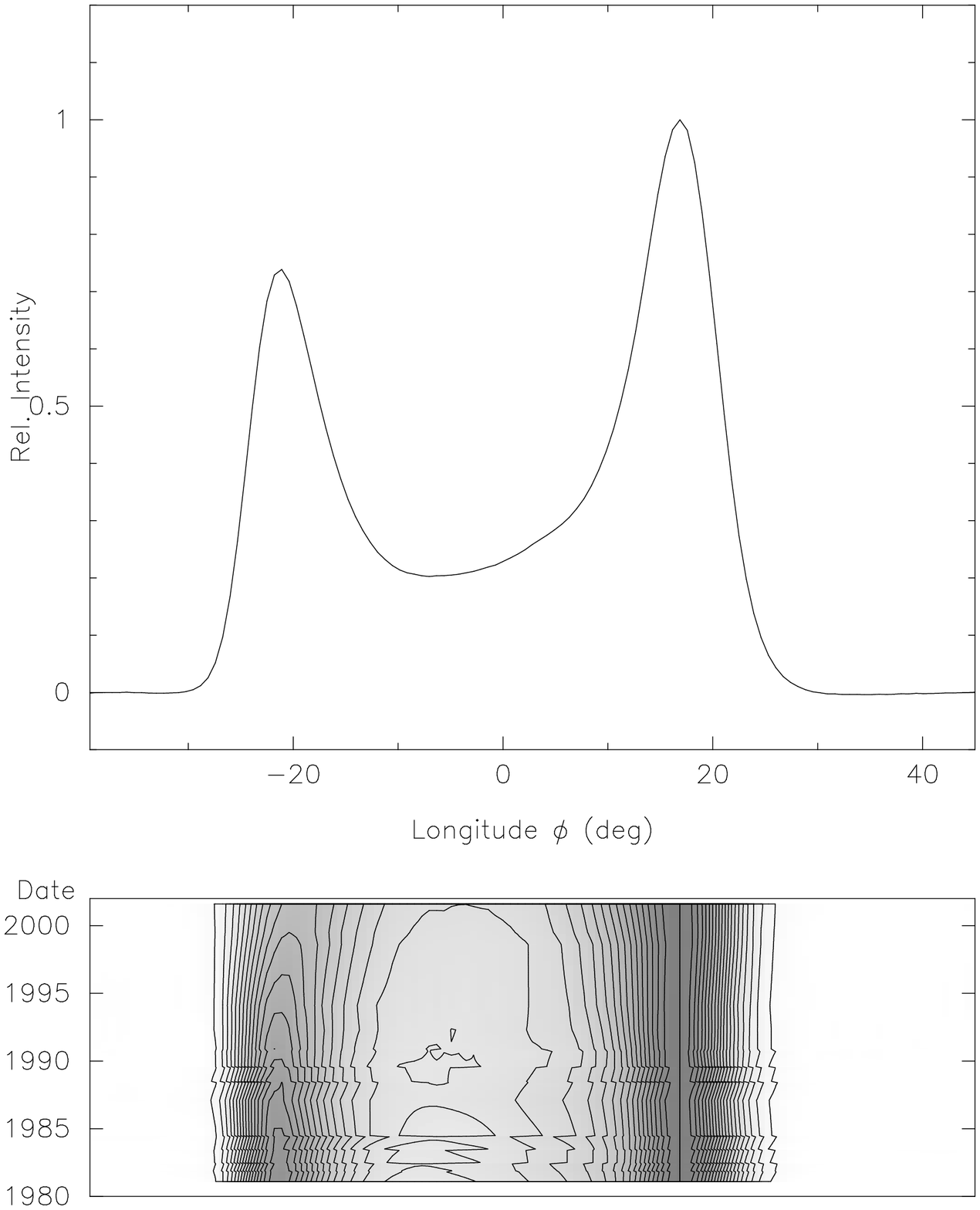}
\caption{PSR B1913+16 Pulse Profile.  Top: time--averaged profile.  
Bottom:
Equal--intensity contour plot, as a function of longitude and time.  
The right peak normalized to constant (unity)
intensity, and set at fixed longitude.  Note that the left
peak of the profile declines  in intensity and approaches the 
right peak while the central saddle fills in and the outer 
skirts of the profile barely change
through the timespan of observations.
\label{fig:rawprof}}
\end{figure*}

Fig.~\ref{fig:rawprof} shows the resulting profiles as a function of time, 
displayed as a contour plot with the right peak arbitrarily normalized to 
constant longitude and  intensity.  There are several notable items revealed 
in the Fig.~\ref{fig:rawprof} contour plot.  First, the leading-- to 
trailing--component peak 
intensity ratio $(I_{leading}/I_{trailing})$ continued its secular decline 
of $\sim 1\% /$ yr, which \citet{Weisberg89} attributed to our line of sight 
slowly precessing across a patchy beam. In addition, the peaks have
clearly been moving together with time in the last decade, as first 
announced 
by  \citet{Kramer98}.  Furthermore, it is evident that the saddle is 
filling in 
with time.  Surprisingly, however, the separation and shape of much of the  
profile outside the peaks remain virtually unchanged over the twenty-year 
timespan of data.  While precession across a circular, hollow--cone beam 
would naturally be expected to result in more pronounced changes on its 
inner portions where the radius of curvature is small than on its outer 
segments,  we will show below that our data cannot be explained in this 
fashion alone.  Instead we will find from model fits that the beam must 
be elongated in the meridional (i.e., north--south) direction, and 
hourglass--shaped.

\section{Analysis}

\begin{figure*}
\plotone{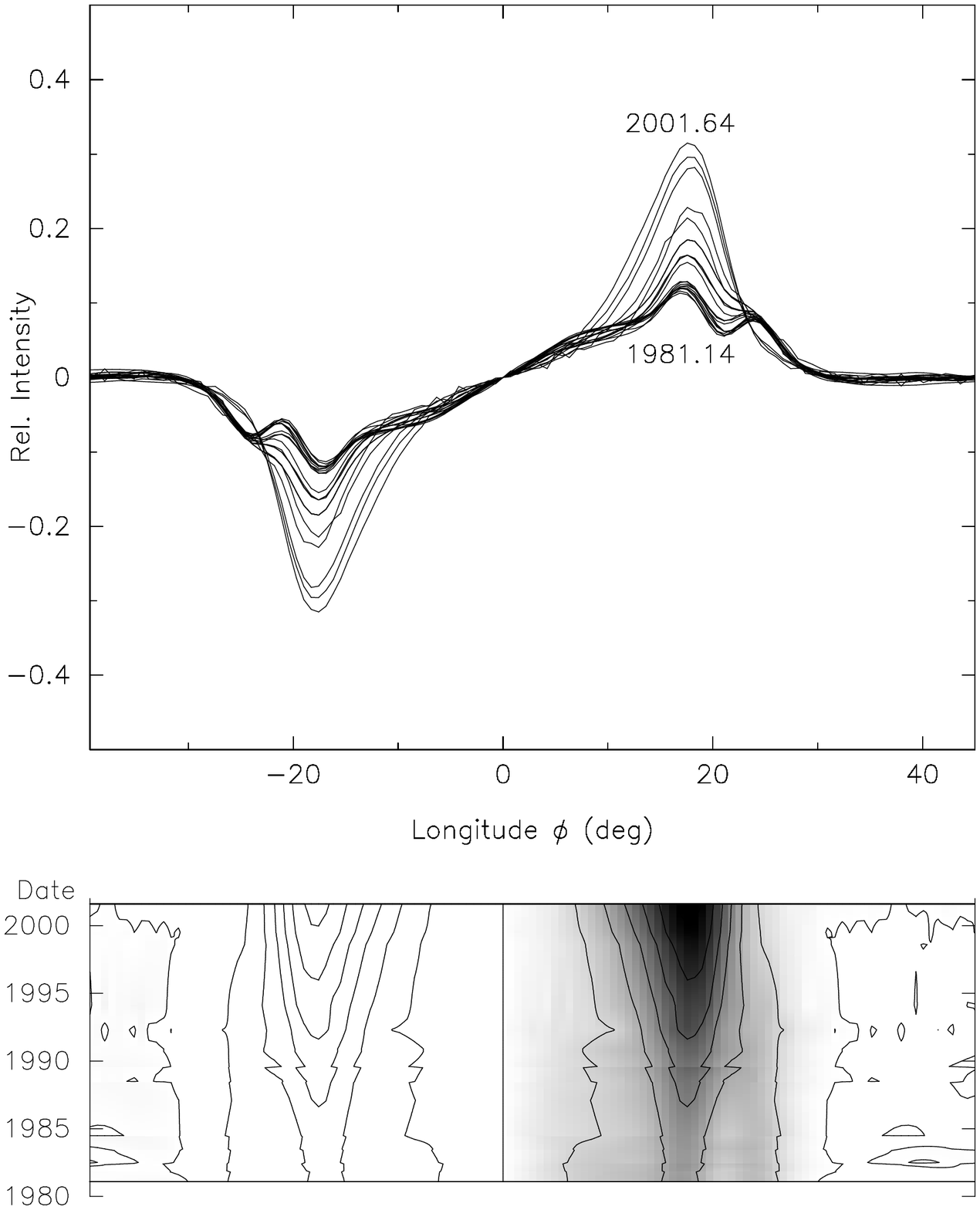}
\caption{Odd Component of PSR B1913+16 Pulse Profile. Top: Individual session 
profiles.  Bottom: Equal--intensity contour plot, as a function of longitude 
and time.
\label{fig:oddprof}}
\end{figure*}

\begin{figure*}
\epsscale{0.8}
\plotone{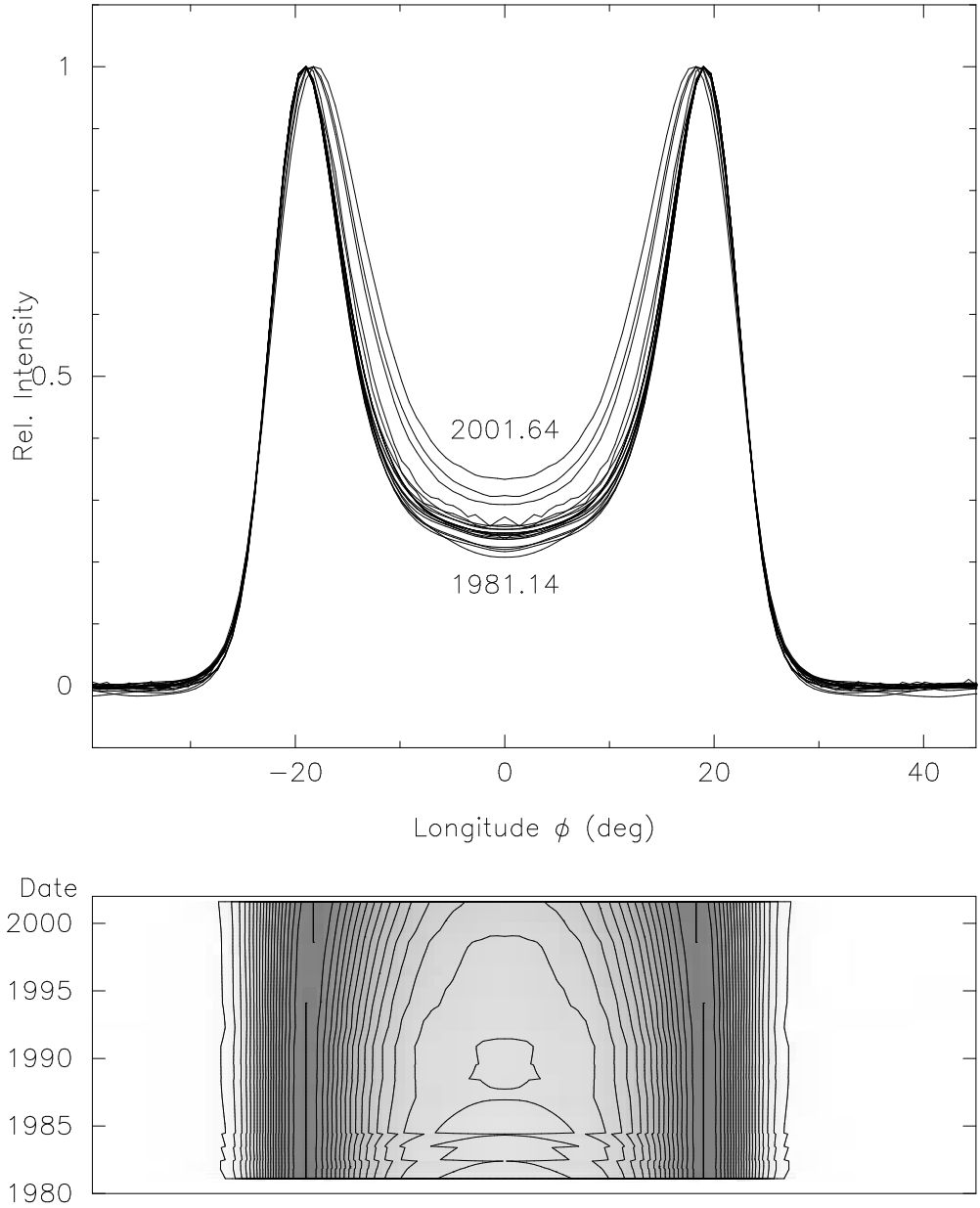}
\caption{Even Component of PSR B1913+16 Pulse Profile. Top: time--averaged 
profile.  Bottom: Equal--intensity contour plot, as a function of longitude 
and time.  As in Fig.~\ref{fig:rawprof}, note that the
profile peak separation decreases while the central saddle fills in and
the outer  skirts remain virtually unchanged or even spread slightly through 
the observing timespan.  
\label{fig:evenprof}}
\end{figure*}

Following up on our finding that our line of sight is slowly precessing 
across a patchy beam,
we decomposed each epoch's profile into an odd and an even part, 
associating the former with the beam patchiness and the latter with 
the overall hollow--cone beam structure. Figs.~\ref{fig:oddprof} and 
\ref{fig:evenprof} display the odd and even components, respectively.  
All subsequent global beam analysis was performed on the even parts alone.  
The pulsar beamwidth $w$ at various  intensity levels was 
determined from 
these even profiles, providing the raw data for the analysis (see below).

\subsection{The Beam Model}

 In our model, 
the beam half--width $\rho_j$ for a given $j-th$ constant intensity  level 
is a function of scale factor $s$ (of order unity); fiducial half--width 
$\rho_{0,j}$; elongation parameters $R$ and $a$; and north--south (i.e.,
meridional) offset  from beam axis $t$; as follows:

\begin{equation}
\rho_j = s\rho_{0,j} (1 + \frac{(R-1) t^2} {a^2+t^2}).
\end{equation}

Of the two elongation parameters, $R$ is essentially the axial ratio 
of the elongated beam, i.e., approximately the ratio of its 
north--south to east--west extent; while $a$ gives a 
$t$--dependent adjustment on the shape.

\setcounter{footnote}{0}

\subsection{The Geometrical Model}

Eq. 1 alone is not sufficient to relate our measured $w$'s in longitude
units to our beam model, because we do not know the full observing geometry
{\it{a priori}}.  For example, the angle of the beam axis with 
respect to the observer, to the spin axis, and to the orbital axis are 
unknown. [Note that the {\it{sine}} of the orbital inclination, $\sin i$, 
is already 
known from pulse timing measurements \citep{TW89}, as discussed below.]

\begin{figure*}
\plotone{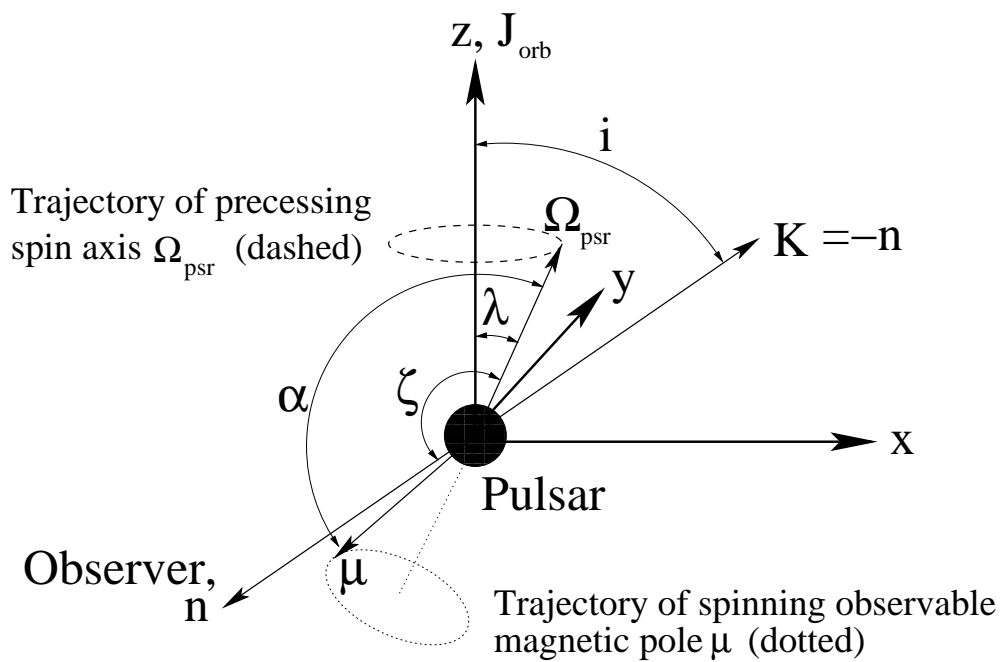}
\caption{Precession geometry.  The dashed ellipse shows the trajectory of the
pulsar spin angular momentum vector $\hat{\Omega}_{psr}(t)$  resulting from 
geodetic precession.  The vector is shown for epoch $T_0$.
The pulsar--spin--induced trajectory of the observable magnetic pole 
$\hat{\mu}$ about $-\hat{\Omega}_{psr}(t)$ is shown, also for epoch $T_0$, as 
a dotted ellipse.  All angles correspond qualitatively to those we find through
model fitting (see text).
\label{fig:geom}}
\end{figure*}

Consequently we have to fit simultaneously for several additional 
geometrical quantities in addition to the elongated beam quantities 
delineated in Eq. 1.
In what follows we will use the work of \citet{Kramer98} as a general guide,
but we will adhere to a right--handed coordinate system notation as in
\citet{DT92} and \citet{EW01}. (See Fig.~\ref{fig:geom}.)  The inclination
$i$ is the angle between
the orbital angular momentum vector $\vec{J}_{orb}$; and the line of sight 
vector 
from the observer to the pulsar, $\hat{K}$.  The spin--orbit misalignment 
angle
$\lambda$ is measured between  $\vec{J}_{orb}$ and the pulsar spin
angular momentum vector  $\hat{\Omega}_{psr}(t)$, also called the 
north spin pole.\footnote{Our angles
$i$ and $\lambda$ are operationally the same as Kramer's,
although he actually defines each as the angle between two vectors, each of 
which points exactly oppositely to the two that we use.  The $\lambda$ of
\citet{DT92} represents a different angle.} Geodetic precession of 
$\hat{\Omega}_{psr}(t)$ about $\vec{J}_{orb}$ occurs
in the $\vec{J}_{orb} \times \hat{\Omega}_{psr}(t)$ direction at a 
rate of $\Omega_{prec}=1.213\deg /$ yr
[See \citet{BarkO75}; \citet{Weisberg89} for the formula].  The
emission beam colatitude $\alpha$ is the angle between 
$\hat{\Omega}_{psr}(t)$ and
the visible magnetic axis $\hat{\mu}$,\footnote{Kramer's $\alpha$ 
is the supplement of ours since he measures it with respect to his 
$\Omega$, which points oppositely to our pulsar spin angular 
momentum vector $\hat{\Omega}_{psr}(t)$} while line--of--sight 
colatitude  $\zeta(t)$
is the angle between $\hat{\Omega}_{psr}(t)$ and the line of 
sight from the pulsar
to observer $\hat{n}$, where $\hat{n} = - \hat{K}$. 
\footnote{ \citet{Kramer98} defines an angle similar to our 
$\zeta(t)$, which he calls $\beta(t)$. His $\beta(t)$, which 
is measured from the {\it{negative}} pulsar spin angular momentum 
axis, is the supplement of our $\zeta(t)$. Note that our angle 
$\beta$ defined below is a different angle from Kramer's $\beta$.}

Now set up a right--handed coordinate system $(\hat{x},\hat{y},
\hat{z})$ where $\hat{z}$ points
along the orbital angular momentum $\vec{J}_{orb}$; and the
$\hat{x}$--vector points such that the plane defined by the 
$\hat{x}$-- and $\hat{z}$-- vectors contains the
observer--pulsar line--of--sight vector $\hat{K}$ (with 
$\hat{x}$ pointing
more toward $\hat{K}$ than toward its opposite, $\hat{n}$).  
The reference time $T_0$ is defined as the time when the  
(precessing) spin angular momentum vector
$\hat{\Omega}_{psr}(t)$ lies in this same plane containing 
$\hat{x}$ and $\hat{z}$ (and $\hat{K}$) and points most directly 
toward $\hat{x}$ and away 
from the observer.\footnote{The \citet{Kramer98} $T_0$ is the same 
as ours. It is defined to be the time when his pulsar spin axis, 
$\Omega$ (which points {\it{oppositely}} to our pulsar spin  
angular momentum vector $\hat{\Omega}_{psr}(t)$), points most 
directly {\it{toward}} the observer.} We can then define a 
precessional phase 
$\phi_{prec}(t) = \Omega_{prec}(t-T_0)$.\footnote{The 
\citet{Kramer98} equation 
for precessional phase differs by a sign,  because it
describes the time evolution of his vector $\Omega$ pointing 
{\it{oppositely}} to our
pulsar spin  angular momentum vector $\hat{\Omega}_{psr}(t)$, 
in a
coordinate system whose $\hat{y}$ and $\hat{z}$ axes are inverted 
with respect 
to ours.}  The position of the precessing spin axis, 
$\hat{\Omega}_{psr}(t)$, 
can be resolved onto the Cartesian coordinate system as follows:

\begin{equation}
\hat{\Omega}_{psr}(t) = 
\hat{x} \cos \phi_{prec}(t) \sin \lambda + 
\hat{y} \sin \phi_{prec}(t) \sin \lambda + \hat{z} \cos \lambda;
\end{equation}

while $\hat{K}$ is resolved onto the same system as

\begin{equation}
\hat{K} =  \hat{x} \sin i +  \hat{z} \cos i.
\end{equation}

The observer's line of sight colatitude $\zeta(t)$  can then be 
found at any
epoch because the definition of the dot product requires that

\begin{equation}
\cos \zeta(t) = 
\hat{\Omega}_{psr}(t) \cdot \hat{n} = \hat{\Omega}_{psr}(t) \cdot (-\hat{K}),
\end{equation}

so that we have

\begin{equation}
\cos \zeta(t) = - \cos \phi_{prec}(t) \sin \lambda \sin i - \cos \lambda \cos i.
\end{equation}

Finally, $\beta(t)$, the impact parameter of the pulsar--observer line of sight 
$\hat{n}$ with respect to the visible magnetic axis $\hat{\mu}$,   is given by
\footnote{See \citet{EW01} for a discussion of impact parameter $\beta$ in the 
context of emission beam models.  Kramer's impact parameter $\sigma$ differs
from our angle $\beta(t)$ by a sign since the two angles in the equation are
the supplements of Kramer's.}

\begin{equation}
\beta(t) = \zeta(t)- \alpha.
\end{equation}

\subsection{Beam / Geometrical Model Fits}

For our quantitative beam modelling, we first determined the pulse width $w$ 
at fourteen intensity levels $I_j$ for each of twenty-two epochs in the even 
profiles, and  we then fitted 
 to these data the elongated, north--south symmetric beam model of Eq. 1 plus 
the geometrical model delineated in the last section.  The fourteen $\rho_{0,j}$
were fixed at values determined iteratively to optimize the fits described
below. 
\footnote{The $\rho_{0,j}$ are an arbitrary set that is associated with the
$j$ (arbitrarily) selected intensity levels.} 
The orbital inclination $i$ was fixed at a
value of $47\fdg20$ or its supplement, as determined from pulse timing
measurements of $\sin i$.  The quantities
$\alpha,\lambda,  T_0, s, R,$ and  $a,$  were then determined from a weighted 
least--squares fitting procedure. \citet{Kramer98} also fitted for quantities
similar to the first three and four of these parameters with his azimuthally 
symmetric beam model and data on the separation between the profile 
{\em{peaks}} only.  We are able to fit for two additional parameters
quantifying the shape of the beam because of our measurements of beamwidth
at many different intensity levels, rather than only at the profile peaks. 

\begin{figure*}
\plotone{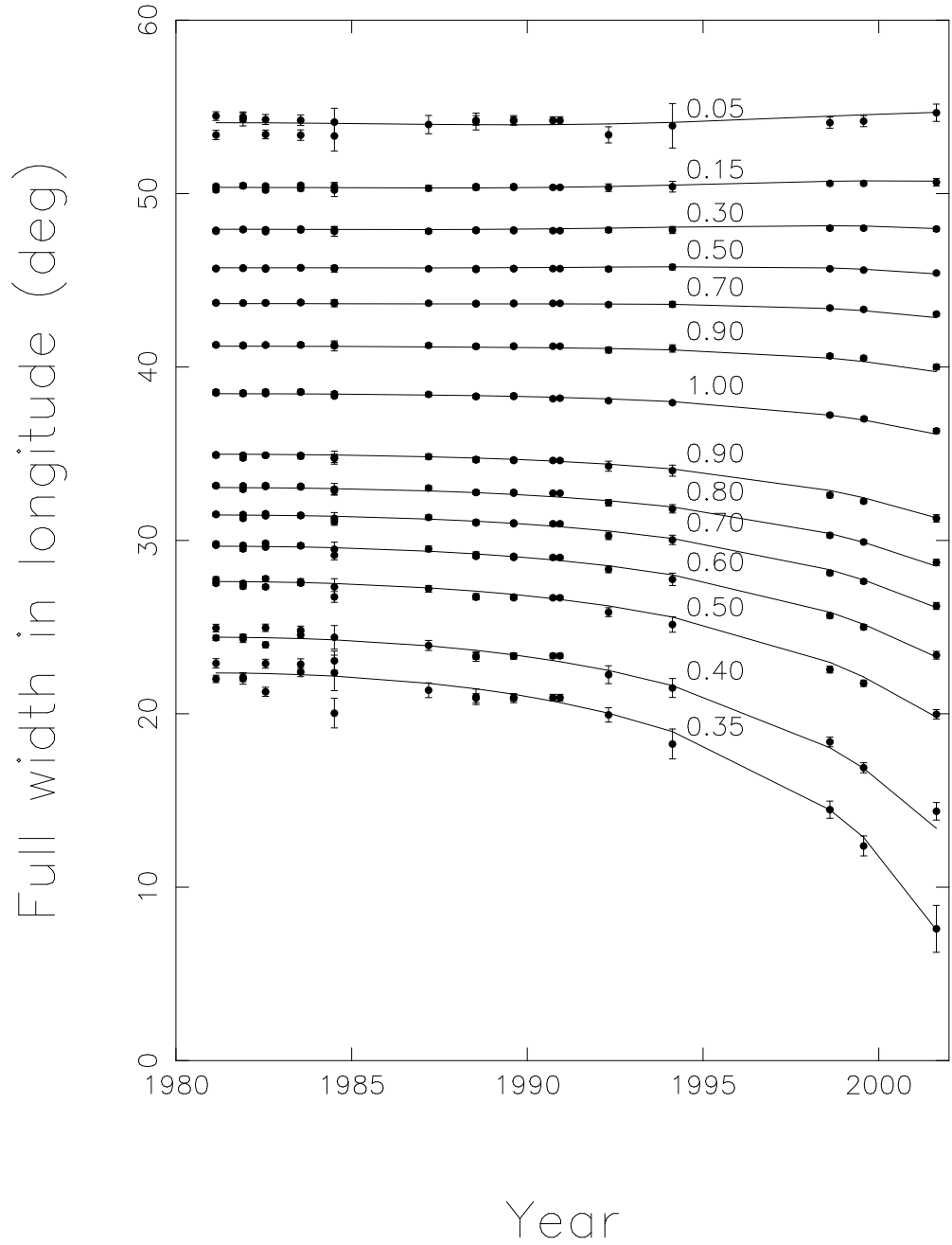}
\caption{Even component pulsewidth versus time at the fourteen relative 
intensity levels listed on the contours.
The points and error bars represent our data (also shown as contours at 
the bottom of Fig.~\ref{fig:evenprof}); the curves display the model
fitted to these data.
\label{fig:data&fit}}
\end{figure*}

As also found  by \citet{Kramer98}, there are four equivalent beam solutions:
Two of them result from the ambiguity in $i$ discussed above, based on the
fact that pulse timing yields $\sin i$. Then each of these two solutions is
also degenerate in that $(\alpha,\lambda,T_0)$ and $(\pi-\alpha, \pi-\lambda, 
 T_0 - \pi / \Omega_{prec})$ are indistinguishable.  Fig.~\ref{fig:data&fit} 
displays the measured pulse 
widths, and the contours of constant intensity  resulting from our final 
best fits to these data.  (All four of the degenerate fits yield the same
results on this figure, which is why they are degenerate.)

\begin{figure*}
\plottwo{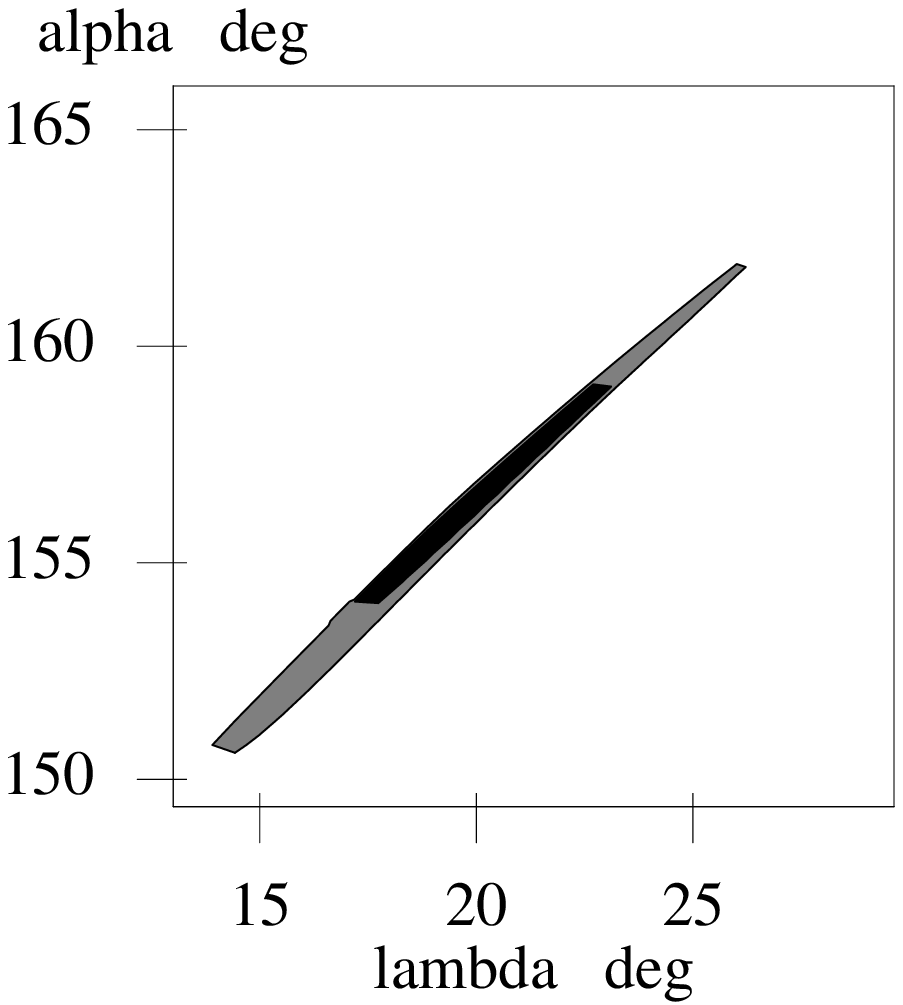} {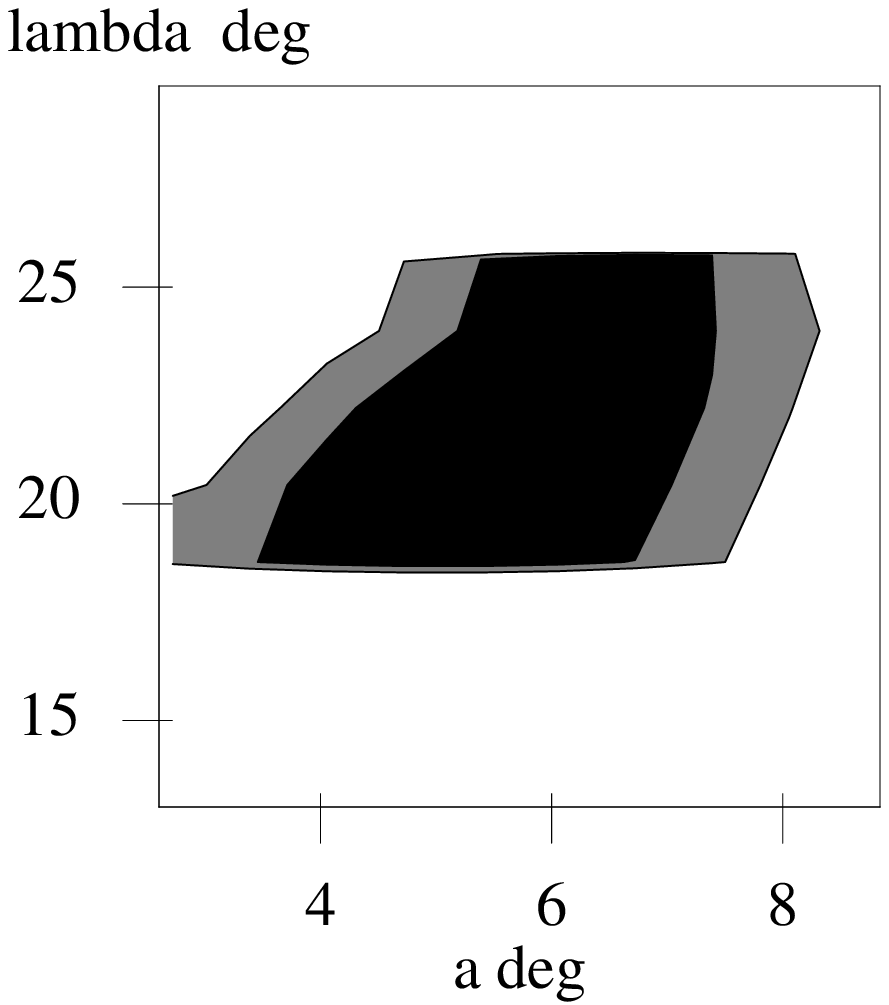}
\caption{Contours of equal $\Delta{\chi}^2$, holding
two parameters fixed at the  values indicated on each graph and allowing all 
other parameters to be fitted. In both graphs, the inner contour is at
the $\Delta{\chi}^2=1$ level; the outer is at $\Delta{\chi}^2=2$
above the best fit value.
\label{fig:chisq}}
\end{figure*}

Fortunately we can rule out the two
high--$\lambda$ solutions because of evolutionary considerations:  The
spin and orbital angular momentum vectors were almost certainly aligned
due to mass exchange before the last supernova occurred, and asymmetries
in that explosion that would misalign the two vectors are constrained by
the observed pulsar velocity distribution (Bailes 1988; Kramer 1998;
Wex et al 2000).  Table 1 records quantitatively the two surviving
low--$\lambda$ fits.  Because some of the fitted parameters are
highly covariant, the formal errors must be read with caution.  In order to
further examine the  uncertainties in fitted quantities, we explored 
$\chi^2$--space
two parameters at a time in the following fashion.  We  stepped the two chosen
parameters through a grid of values in the vicinity of the $\chi^2$--minimum,
holding them fixed at each grid point while allowing all other parameters
to be fitted.  The resulting $\chi^2$ at each such grid point was then 
calculated 
and a contour map of $\chi^2$ as a function of the two parameters was
generated.  We display the two maps showing the most serious covariance in
Fig.~\ref{fig:chisq}.  Note especially that $\chi^2$ contours associated with 
$\alpha$ and $\lambda$ indicate that the possible range of their errors is 
much larger than indicated in our formal uncertainties.

\subsection{Polarization Data and the Rotating Vector Model}

\begin{figure*}
\plottwo{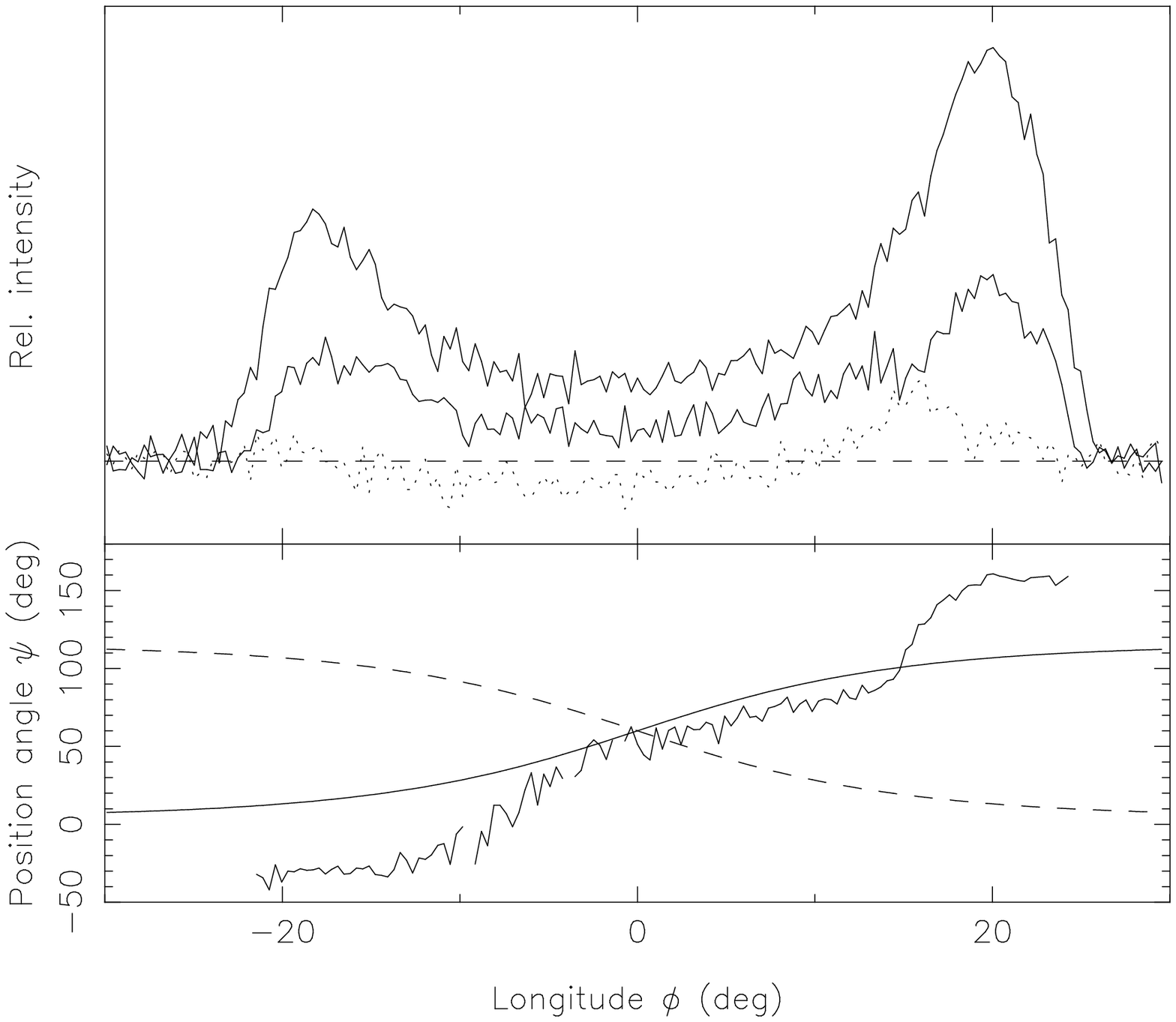}{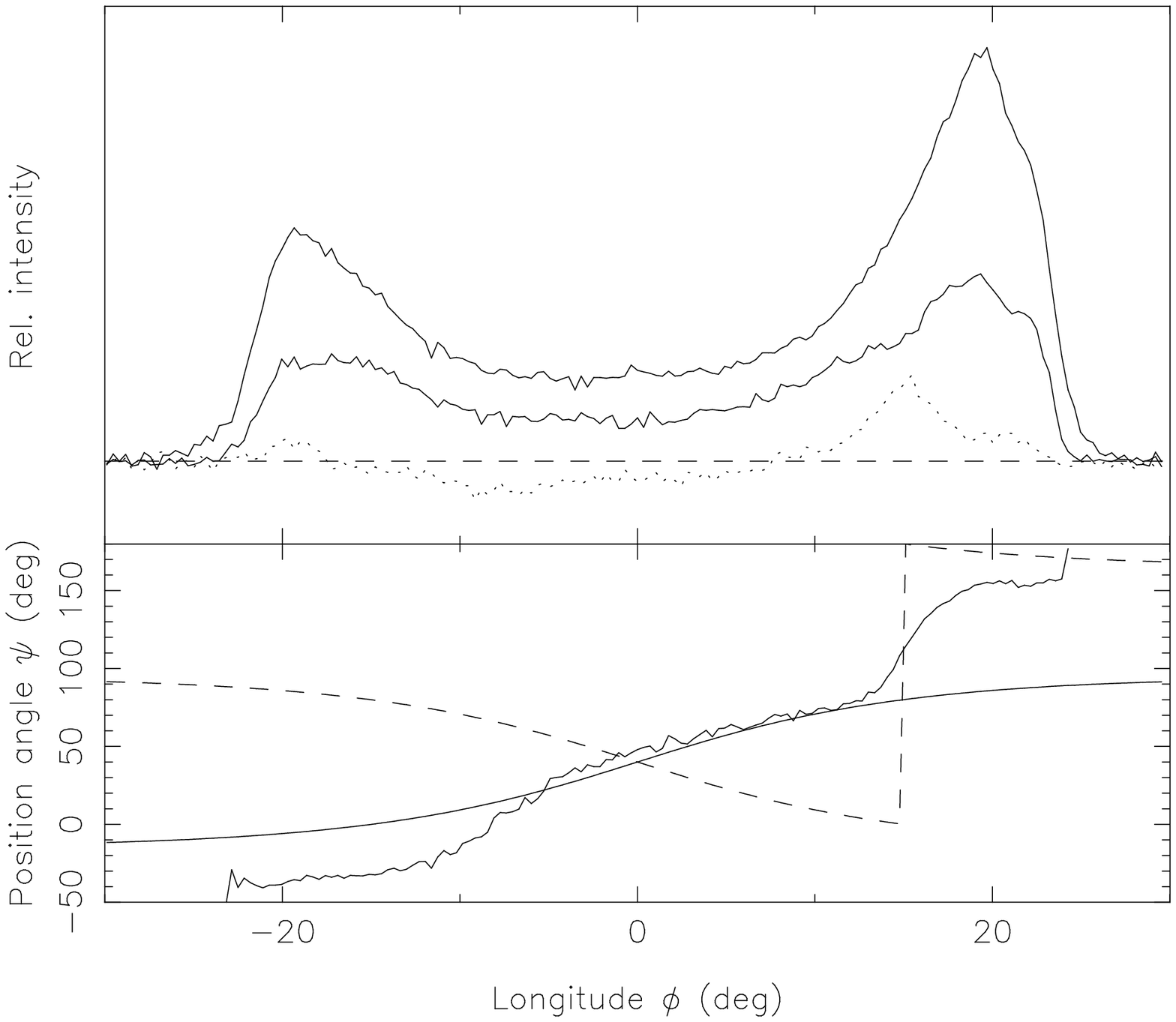}
\caption{Polarization profiles at 1403 MHz in 1998 (left), and 1999 (right).
In each plot, the top panel shows the polarized profile, with total power 
Stokes parameter $I$  the highest profile curve; linear power $L$ the next 
highest; and  circularly polarized  Stokes parameter $V$   dotted. The bottom 
panel displays the  position angle as observed and as expected from applying 
the two low--$\lambda$ beam -- geometrical  fits listed in Table 1 to the
rotating vector  $(RVM)$ model of polarization.  Solution 1 yields  
the smooth solid curve, while solution 2 leads to the smooth dashed curve. 
Each model was derived from beamshape data independent of polarization.
For each epoch, one model conforms at least crudely to the position angle 
data while one 
is grossly wrong even in the sign of its slope.  Note:  The  position angle 
$\psi$  conforms to the observers' convention, increasing counterclockwise 
on the sky. The RVM model is cast in terms of a position angle
$\psi'(=-\psi)$ that increases clockwise on the sky \citep{EW01}. 
\label{fig:stokes}}
\end{figure*}

In an effort to provide additional information on the emission
beam geometry, we  have also measured the polarization properties of the pulse 
profile 
starting with our 1998 observing session,  using the Princeton Mark IV backend \citep{Stairs00}. Early pulsar observations revealed a characteristic
$S$--shaped sweep of polarization position angle across many pulse profiles.
According to the rotating vector model (RVM) of
\citet{RC69}, this signature can be understood
as the projection of the rotating neutron star's magnetic field lines onto the
observer's line of sight, because the emission is
locally polarized in (or perpendicular to) the plane of the field lines. In 
theory, it is 
possible to fit linear polarization position angle data to the RVM in order
to derive the emission beam geometry. In practice, however, it has been
shown that unambiguous geometrical solutions from RVM fits to linear 
polarization data are rare, for two principal reasons \citep{EW01}:
First, the observable emission from most pulsars occupies such a narrow
longitude range that RVM fit parameters are highly covariant; and second,
many pulsars' emission does not conform simply to the RVM itself.
Our polarized profiles (Fig.~\ref{fig:stokes}), indeed exhibit a complicated 
form.  \citet{CWB90} also measured and remarked on the complicated nature 
of the position angle curve.  They noted
that its overall sweep of $\sim 195 \degr$ is not allowed in the rotating 
vector model, and that the rapid swings indicate switching between orthogonal
emission modes.  These features have prevented us from being able to  perform  
RVM fits to the position angle data that yield unique values for geometric
parameters.

Despite these complications, the position angle data
do provide critical information that enables us to select one and reject 
one of 
our two remaining beam model solutions. To do so, we calculated 
position angle curves  for the best--fitting parameter values
of both solutions listed in Table 1.  For each solution, we first
calculated the  observers' line of sight colatitude for the observing 
epoch $\zeta(t)$, from Eq. 5.  We then determined the  expected RVM linearly 
polarized  position angle \footnote{Note that  position angle $\psi$ follows 
the observers'
convention, increasing counterclockwise on the sky.}  $\psi$ with 
longitude $\phi$ curve from the classic RVM equation [corrected for signs; 
see  \citet{EW01}]:

\begin{equation}
\tan (\psi - \psi_0) = -\frac{\sin \alpha \sin (\phi - \phi_0)}
{\sin \zeta \cos \alpha - \cos \zeta \sin \alpha \cos (\phi - \phi_0)}.
\end{equation}
Note that  $\alpha$ and $\zeta(t)$ go to their supplements in the alternate
solutions, which also causes the impact parameter $\beta(t)$ to flip sign 
(see Eq. 6). This
has crucial implications for the RVM and the polarized profile,
since the slope of the polarized linear position angle curve, 
$d\psi / d\phi = - \sin \alpha / \sin \beta$ at the center of the pulse, 
also changes sign because
our line of sight moves from one side of the magnetic axis to the other
[see \citet{EW01} for details].

The results for both solutions are shown in Fig.~\ref{fig:stokes}  atop the 
data.  As expected, one
model (Solution 2, with $\alpha\sim 22\degr; \zeta\sim29\degr; 
\beta\sim6\degr$) clearly gives a grossly wrong position angle sweep with the 
wrong sign. The complementary model,  (Solution 1, $\alpha\sim158\degr;  
\zeta\sim151\degr; \beta\sim-6\degr$) sweeps
in position angle in the same sense as the data.

It appears especially heartening that Solution  1 matches the observed position 
angles in the central regions of the pulse, not just in the sign of the slope 
but also in a more detailed fashion.  As a result, we closely studied the 
position angle measurements in our data as well as the earlier polarimetry of 
Cordes et al  (1990).  The precession should
cause the position angle curve to vary with time as $\zeta$  changes.
The predicted changes are rather small over the timespan of our Mark IV data. 
 However, we were disappointed to note that the Cordes
et al. data from epoch 1988.5, which should show a
position angle curve measurably different from current data, do {\em{not}}.
Rather, the much earlier Cordes et al. position angle curve is essentially 
indistinguishable from our Mark IV curves to within the noise. 
We conclude that the pulsar's emission is not following the RVM in detail.
Nevertheless,  it seems likely that the  slope of the position angle
curve near the symmetry axis of the pulse should still be related to the
apparent rotation of the magnetic field lines emanating from the magnetic 
pole as they sweep across our line of sight, 
and hence successfully chooses the sign of $\beta$ and thus Solution 1.

\section{Discussion}
\begin{figure*}
\plotone{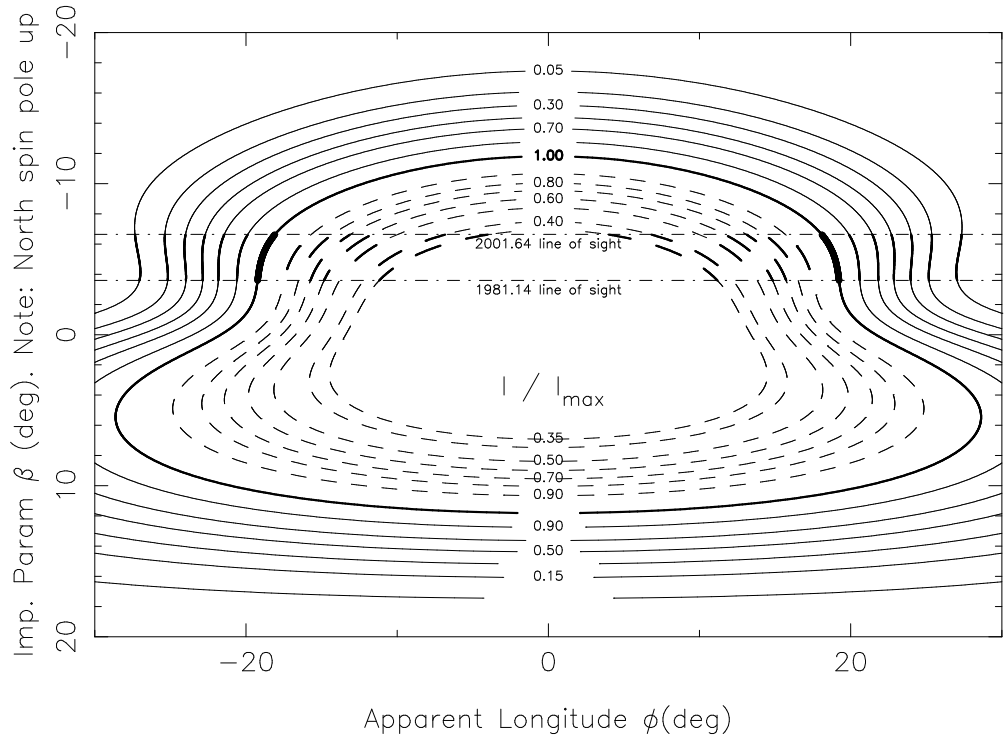}
\caption{The best--fitting, even component beam model, plotted as a function
of longitude and north--south angular offset from the beam center. (Longitude
is a direct measurable, since $360\degr = 1$ pulsar rotation.) The peak 
$(1.0)$ intensity contour is bold, and decreasing contours inside it are dashed.
The apparent north--south beam 
asymmetry is due to the convergence of meridians (lines of constant longitude) 
toward the pole.  See Fig.~\ref{fig:beamarc} for a view of the model in 
undistorted 
(great--circle arc) space.  Currently we have directly sampled only those 
portions of the beam lying between the two horizontal lines.  Note that this
region of the plot is also displayed in the 
bottom of Fig.~\ref{fig:evenprof} and in Fig.~\ref{fig:data&fit}. (The
earlier figures show  pulsewidth as a function of {\it{time}}, whereas we have 
used our fitted geometrical parameters to transform the time axis into an angular 
north--south offset on the pulsar in the above figure.)
\label{fig:beamlong}}
\end{figure*}

The fitted equal--intensity contours alone, without
the data, are shown in Fig.~\ref{fig:beamlong}, over a much larger 
north--south range than we have directly observed.  We delineate the 
portion of the beam that we have directly sampled over the past twenty years 
with two horizontal lines.  Note that we have observed only a small part of 
the beam in the vertical direction.  Consequently our model is by no 
means unique, and it is probably incorrect in its details.  Most notably,
our assumption of north--south beam symmetry is currently untestable since all
observations probe only the northern side of the beam.
Nevertheless it is clear that {\it{any}} successful beam model must be able to
match the striking features shown in the data and in our model:  a shrinkage
with time of the inner and peak equal--intensity contours, combined with
uniform  or even slightly growing outer contour widths.  

\begin{figure*}
\plotone{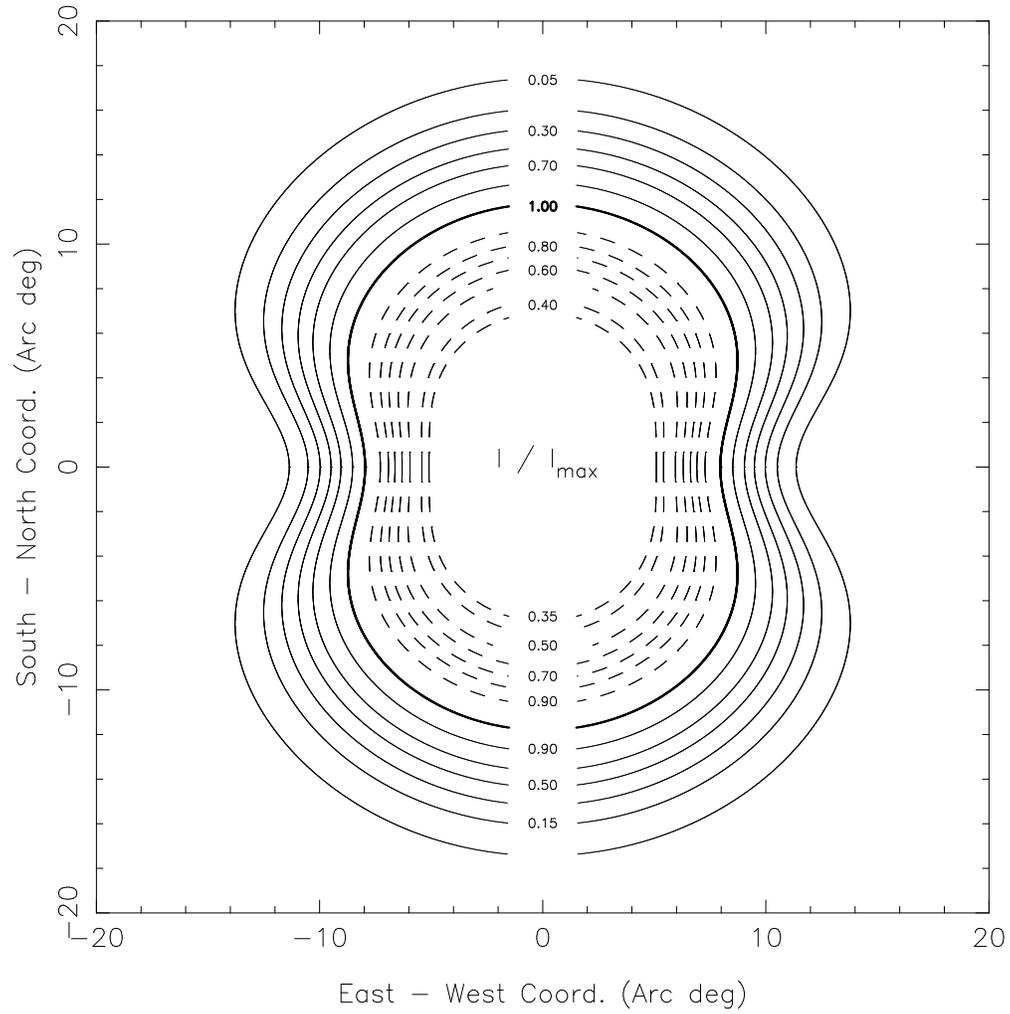}
\caption{The best--fitting, even component beam model, plotted as a function
of east--west and north--south angular offset from the beam center.  The peak 
$(1.0)$ intensity contour is bold, and decreasing contours inside it are dashed.
\label{fig:beamarc}}.
\end{figure*}

The apparent north-south asymmetry in Fig.~\ref{fig:beamlong} is an artefact
of displaying the horizontal axis in directly measurable longitude units 
(where $360 \degr = 1$ pulsar
rotation), which are portions of small circles whose arc length varies as one
approaches a pole.  Fig.~\ref{fig:beamarc} displays the best fitting beam model
in true great--circle arc units on both axes, and the expected  north--south 
(and east--west) symmetries are evident in this display.  It is surprising 
that the beam takes on an "hourglass" shape in order to match our observations 
of pulsewidth that varies strongly at small radii and remains virtually 
unchanged or even growing slightly at larger radii.  It is interesting to 
note that \citet{Link01} have also derived an hourglass--shaped beam for 
PSR B1828-11, an isolated pulsar that is also undergoing spin axis precession.
Thus the only two pulsars whose beams have been mapped in the meridional
direction show qualitatively similar beamshapes.

Others have investigated pulsar beamshapes. Most studies are based  on 
rather indirect observational arguments, usually involving analyses of
position angle and pulsewidth data in the pulsar population.  \citet{J80}
and \citet{NV83}  found  significant north--south beam elongation.  However, 
\citet{LM88} and \citet{B98} 
indicated that the emission is consistent with a circular cone.  \citet{MD99} 
also supported circular cones although finding marginal evidence for meridional {\it{compression}}.  \citet{B90} 
and  \citet{M93} also point out that the beam will be meridionally compressed  
if its boundary is defined by the last closed field lines
in a spinning  neutron star with a tilted magnetic axis.   

There is clearly no consensus on the overall beam elongation.  In addition, 
we are not aware
of any observational or theoretical studies indicating the hourglass shape,
other than ours and \citet{Link01}.

\section{Conclusion}
We have used twenty years of pulse profile measurements on PSR B1913+16 to 
detect geodetic precession of the spin axis and to model the emission beam 
geometry. Evolutionary considerations and polarization measurements enable
us to choose a single preferred model from four equivalent solutions.  While
we have only sampled a limited portion of the beam to date, our model indicates
that the beam is elongated and hourglass--shaped in the latitude direction.
Future observations will enable us to study the two dimensional beam structure 
at larger impact parameters, further constraining the details of the beamshape.

\acknowledgements{We thank M. Kramer, N. Wex, and A. Karastergiou for thoughtful
discussions.  J. Bell Burnell, A. de la Fuente, and T. Green helped with some
of the observations.   JMW and JHT have been supported by grants from
the National Science Foundation, the latest being AST 0098540 and AST 9618357.  Arecibo Observatory is operated by 
Cornell University under cooperative agreement with the NSF.}

\newpage
\clearpage

\clearpage

\clearpage

\clearpage

\clearpage

\clearpage

\clearpage

\clearpage

\clearpage

\clearpage
\begin{deluxetable}{llll}
\tabletypesize{\normalsize}
\tablewidth{0pt}
\tablenum{1}
\tablecolumns{4}
\tablecaption{Preferred Low--$\lambda$ Beam and Geometrical Model Solutions.
\label{tbl:1}}
\tablehead
{Para- 	& Description 			  & Solution 1	& Solution 2		\\
 meter	&				  &		&	 		\\} 
\startdata
$i$	  & Orbital inclination 	  & $47\fdg2$	& $132\fdg8$(fixed) 	\\
$\alpha$  & Emission beam colatitude	  & $157\fdg5$	& $22\fdg5 \pm0\fdg3$ 	\\ 
$\lambda$ & Spin--orbit misalignment angle& $21\fdg1$	& $21\fdg1 \pm0\fdg3$	\\
$T_0$	  & Reference epoch		  & 1981.0	& $2129.4 \pm 0.3$	\\
$s$	  & Scale factor		  & 1.04	& $1.04 \pm 0.01$	\\
$R$	  & Axial ratio			  & 1.59	& $1.59 \pm 0.02$	\\
$a$	  & Shape adjustment		  & $5\fdg57$	& $5\fdg57 \pm 0\fdg11$	\\
\enddata
\tablecomments{Fits of the rotating vector model (RVM) to our polarization data
eliminate Solution 2. See text and Fig.~\ref{fig:stokes}.}
\end{deluxetable}

\end{document}